\documentclass[
 reprint,
 superscriptaddress,
 amsmath,amssymb,
 aps,
prl,
]{revtex4-2}

\usepackage{graphicx}
\usepackage{dcolumn}
\usepackage{bm}
\usepackage{hyperref}
\usepackage{siunitx}
\usepackage{xspace}
\usepackage{xcolor}

\newcommand{\ket}[1]{\ensuremath{\lvert #1 \rangle}\xspace}%

\begin{document}

\newcommand{\partitle}[1]{\subsection*{#1}}

\title{Single-qubit detection by collective phase imprinting}


\author{Kritsana Srakaew}
    \affiliation{Munich Center for Quantum Science and Technology (MCQST), 80799 Munich, Germany}
    \affiliation{Max-Planck-Institut f\"{u}r Quantenoptik, 85748 Garching, Germany}

\author{Pascal~Weckesser}
    \affiliation{Munich Center for Quantum Science and Technology (MCQST), 80799 Munich, Germany}
    \affiliation{Max-Planck-Institut f\"{u}r Quantenoptik, 85748 Garching, Germany}

\author{Daniel Adler}   
    \affiliation{Munich Center for Quantum Science and Technology (MCQST), 80799 Munich, Germany}
    \affiliation{Max-Planck-Institut f\"{u}r Quantenoptik, 85748 Garching, Germany}
    
\author{Suchita Agrawal}
    \affiliation{Munich Center for Quantum Science and Technology (MCQST), 80799 Munich, Germany}
    \affiliation{Max-Planck-Institut f\"{u}r Quantenoptik, 85748 Garching, Germany}

\author{David Gröters}
    \affiliation{Munich Center for Quantum Science and Technology (MCQST), 80799 Munich, Germany}
    \affiliation{Max-Planck-Institut f\"{u}r Quantenoptik, 85748 Garching, Germany}
    \affiliation{Fakult\"{a}t f\"{u}r Physik, Ludwig-Maximilians-Universit\"{a}t, 80799 Munich, Germany}

\author{Immanuel Bloch}
    \affiliation{Munich Center for Quantum Science and Technology (MCQST), 80799 Munich, Germany}
    \affiliation{Max-Planck-Institut f\"{u}r Quantenoptik, 85748 Garching, Germany}
    \affiliation{Fakult\"{a}t f\"{u}r Physik, Ludwig-Maximilians-Universit\"{a}t, 80799 Munich, Germany}

\author{Johannes Zeiher}  
    \affiliation{Fakult\"{a}t f\"{u}r Physik, Ludwig-Maximilians-Universit\"{a}t, 80799 Munich, Germany}
    \affiliation{Munich Center for Quantum Science and Technology (MCQST), 80799 Munich, Germany}
    \affiliation{Max-Planck-Institut f\"{u}r Quantenoptik, 85748 Garching, Germany}

\date{\today}

\begin{abstract}

The amplification of quantum information carried by a single quantum excitation is a recurring challenge across diverse quantum platforms.
The coupling between a single qubit and a mesoscopic ensemble of spins, for example, can be leveraged to realize non-destructive detection of the qubit state.
However, realizing robust couplings between such systems is experimentally challenging and typically requires programmable quantum gates or native long-range interactions.
Here, we introduce a platform that couples a single qubit, encoded in the ground-to-Rydberg transition of a control atom, to a Rydberg-dressed target ensemble of ground-state atoms trapped in an optical lattice.
We show that the state of the control qubit can be coherently mapped onto the ensemble via a qubit-controlled collective phase shift.
By Rydberg-dressing the ensemble, the controlled phase shift per target atom becomes independent of the number of target atoms, making the protocol intrinsically insensitive to atom-number fluctuations and atom loss, which are the dominant experimental imperfections in our system.
Exploiting the collective response of up to eight target spins, we demonstrate the efficacy of the scheme by realizing non-destructive detection of a single Rydberg excitation with a state-assignment fidelity of $\mathcal{F} = 99.81^{+0.17}_{-1.47}\,\%$.
Our approach demonstrates the key ingredients for high-fidelity transfer of quantum information from a single qubit to a mesoscopic ensemble, opening a route to non-destructive mid-circuit readout of Rydberg states and to efficient interfaces between single qubits and photonic modes.

\end{abstract}

\maketitle


Amplification of quantum signals requires strong non-linear interactions~\cite{Clerk2010}.
%
%
In the optical domain, parametric amplifiers are a prominent example, where, in an idealized case, an optical non-linearity is leveraged to achieve quantum-limited amplification of coherent few-photon states.
%
A second example is avalanche detectors, where the signal of a single photon is amplified by triggering a stochastic avalanche of electrons in suitable materials.
%
Translating the concept of amplification to qubit or, equivalently, spin systems is possible by engineering a coupling between a single qubit and an ensemble to amplify the qubit state via its mapping to a collective degree of freedom~\cite{Cappellaro2005,Cappellaro2006, Perez2006}.
Importantly, and in contrast to the stochastic amplification found in avalanche detectors, such interactions between quantum degrees of freedom can in principle be unitary.
Applications include the measurement of otherwise inaccessible nuclear spins~\cite{Neumann2010,Taminiau2014} and the use of a central spin as a sensor of its many-body environment to track collective spin dynamics in solid-state systems~\cite{Biswas2025}.\\
Strong, long-range Rydberg interactions provide a natural toolbox to extend these capabilities to neutral-atom systems~\cite{Saffman2010}.
Both single-spin detection and the tracking of many-body dynamics of an interacting ensemble would have immediate applications in Rydberg-based quantum simulators and processors~\cite{Browaeys2020, Menssen2026}.
In previous work, coupling of Rydberg states to neighboring ground-state atoms has been exploited for Rydberg-state detection~\cite{Gunter2012, Ferreira2020, Xu2021} and for the realization of controlled quantum-optical nonlinearities~\cite{Dudin2012, Baur2014, Tiarks2014, Gorniaczyk2014, Srakaew2023, Sumarac2026}.
In these schemes, a single Rydberg excitation controls the optical response of a neighboring ensemble through strong dipolar interactions, switching the system between transmission and absorption.
%
One drawback of these approaches is dephasing within the dense ensembles, which limits the achievable fidelity; another is the relatively large atom number per ensemble required to enter the strongly nonlinear regime~\cite{Goldschmidt2016, Boulier2017}.
Coupling a Rydberg excitation via interactions to a collective spin-degree of freedom offers an alternative approach to Rydberg-state detection.
A possible implementation uses the control of intra- and interspecies interactions in dual-element arrays~\cite{Anand2024} to couple a qubit to its neighbors~\cite{Petrosyan2024, Zhang2025,Vaknin2026}, which can subsequently be read out.
While this strategy is more efficient in qubit usage and avoids dense ensembles, it places stringent demands on the simultaneous maximization of qubit-ensemble interactions and minimization of intra-ensemble interactions, and incurs the experimental overhead of preparing two different atomic species.

\begin{figure*}[t!]
    \centering
    \includegraphics[width=\textwidth]{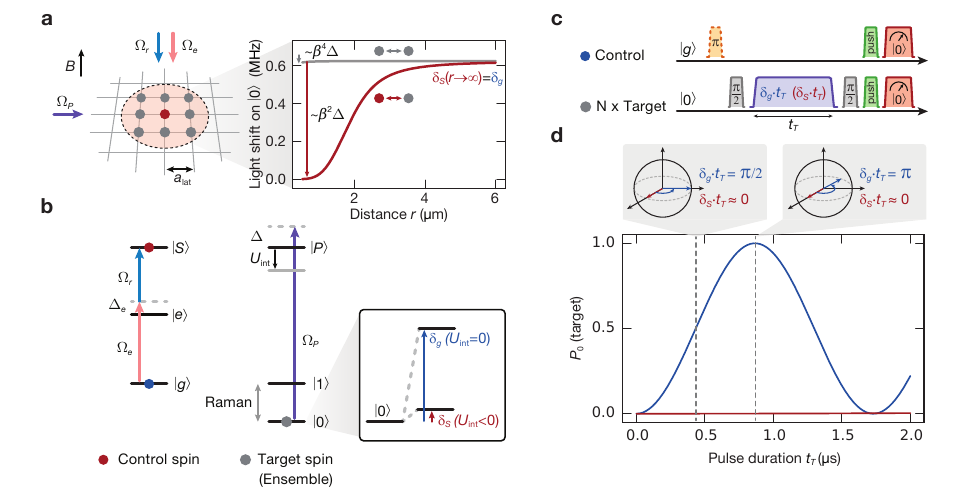}
    \caption{\textbf{Rydberg-mediated light shifts and controlled phase.}
    (\textbf{a}) Schematic of the experimental setup.
    A 3$\times$3 array of $^{87}$Rb atoms is prepared with a lattice spacing of $a_{\mathrm{lat}} = \SI{752}{\nano\meter}$.
    The central control qubit is initialized either in the ground state $\ket{g}$ (blue) or in the Rydberg state $\ket{S}$ (red), while all surrounding target spins (gray) are prepared in the hyperfine state $\ket{0}$.
    The inset shows the distance-dependent AC-Stark shift $\delta_S(r\rightarrow \infty)\sim\beta^2\Delta$ set by the $\ket{S}-\ket{P}$ Rydberg interaction (red curve). On this scale, the fourth-order Rydberg-dressed interaction between neighboring target spins (gray curve) is negligible due to its higher-order scaling $\sim\beta^4\Delta$.
    (\textbf{b}) Electronic level structure and excitation scheme for the control and target spins.
    The control spin is driven from the ground state $\ket{g}$ to the Rydberg state $\ket{S}$ via a two-photon excitation.
    The target spins are off-resonantly coupled to the Rydberg state $\ket{P}$ by a single-photon transition with detuning $\Delta$.
    The transition between $\ket{0}$ and $\ket{1}$ is coupled by a stimulated two-photon Raman process, allowing for fast and high-fidelity population transfer.
    %
    %
    The inset illustrates the AC-Stark shift of the target spins when the control qubit is in the ground state $\ket{g}$ (blue) or the Rydberg state $\ket{S}$ (red).
    (\textbf{c}) Ramsey protocol used to measure the controlled phase imprinted on the target spins.
    After initialization, the control spin is prepared in either the ground ($\ket{g}$) or the Rydberg state ($\ket{S}$), and the phase imprint on the target spins is probed via the residual $\ket{0}$ population $P_0$.
    (\textbf{d}) Expected controlled-phase oscillations for a control–target separation of $r=a_{\mathrm{lat}}$ assuming ideal state preparation.
    When the control spin remains in the ground state, the target spin experiences an AC-Stark shift $\delta_g$.
    In contrast, excitation of the control spin to $\ket{S}$ suppresses the shift, resulting in $\delta_S\approx0$.
    The inset shows the corresponding Bloch-sphere trajectories of the target spin under $\delta_g$ and $\delta_S$.
    }
    \label{fig:1}
\end{figure*}

\begin{figure*}[th!]
    \centering
    \includegraphics[width=\textwidth]{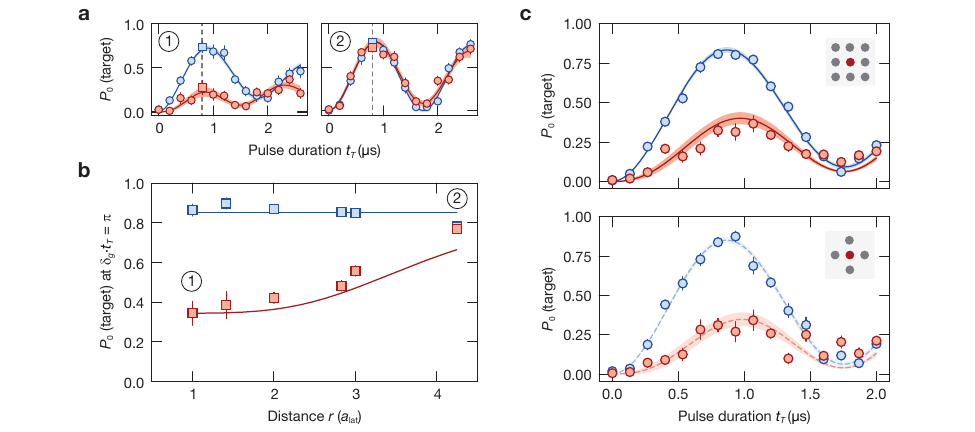}
    \caption{\textbf{Distance-dependence and fluctuation resilience of effective AC-Stark shift.}
    (\textbf{a}) Controlled phase oscillations with the control spin in the ground (blue) or Rydberg (red) state.
    In \textcircled{1}, with the control-target separation set to $r = a_{\mathrm{lat}}$, the Rydberg interaction results in a strong contrast between the two cases. The observed infidelities are primarily limited by state-preparation errors.
    At larger distances ($r=3\sqrt{2} a_{\mathrm{lat}}$) depicted in \textcircled{2}, the Rydberg interaction becomes negligible, resulting in nearly identical phase evolution.
    (\textbf{b}) Distance-dependent phase imprinted for a fixed Ramsey phase.
    Measuring the target-spin population at $\delta_g t_T=\pi$ as a function of control-target distance $r$ is well described by the solid theoretical curve derived from the Rydberg pair potentials, with the state-preparation infidelity included.
    The blue points correspond to reference measurements with the control spin prepared in the ground state.
    (\textbf{c}) Comparison of controlled-phase oscillations for target-spin numbers of $N=8$ (upper) and $N=4$ (lower), where the four diagonal target spins are removed.
    The solid lines in the upper panel show sinusoidal fits, and the shaded regions indicate the corresponding fit uncertainties.
    Treating the oscillation frequency ($\delta_g$), preparation fidelity ($P_{\mathrm{init}}$) and dephasing rate as free fit parameters, we obtain $\delta_g=2\pi\times\SI{578(4)}{\kilo\hertz}$ and $P_{\mathrm{init}}=0.88(13)$.
    The fitted preparation fidelity is in close agreement with an independent estimation of $P_{\mathrm{init}}=0.82(3)$~\cite{SI}.
    In the lower panel, the dashed lines reproduce the fits obtained for $N=8$, illustrating the equivalence of the phase evolution for $N=8$ and $N=4$, consistent with negligible interactions among the target spins.
    Insets depict the corresponding initial state configurations.
    }
    \label{fig:2}
\end{figure*}

Here, we overcome these challenges and demonstrate loss-resilient coupling between a single Rydberg-encoded control qubit and a ground-state target ensemble off-resonantly dressed to a separate Rydberg state.
%
%
Interactions between the resonantly-excited control spin and the Rydberg-dressed target spins modify the AC-Stark shift of each target spin, which we detect as a Ramsey-fringe contrast change.
In this measurement, the collective signal-to-noise ratio grows with the number of target atoms, and the per-target phase shift is independent of the ensemble size, providing intrinsic robustness to atom loss and number fluctuations.
As a first application, we use the scheme to detect a single Rydberg excitation non-destructively with a state-assignment fidelity of $\mathcal{F} = 99.81^{+0.17}_{-1.47}\,\%$.
Our method establishes a route to mid-circuit Rydberg-state detection and points to opportunities for loss-tolerant gates, measurement-based feedback, and many-body sensing in neutral-atom quantum simulators.

The key idea and working mechanism behind our scheme is illustrated in Fig.~\ref{fig:1}.
A single control qubit within a $3\times3$ array of nine atoms is prepared in the ground state \ket{g} and controllably excited to a Rydberg state $\ket{S}$.
The remaining eight target spins are off-resonantly dressed to a neighboring Rydberg $\ket{P}$ state with Rabi frequency $\Omega_P$ and detuning $\Delta$. 
The Rydberg admixture results in a second-order AC-Stark shift $\delta_g = \Omega_P^2/4\Delta$ on the hyperfine ground state $\ket{0}$, while the other hyperfine ground state $\ket{1}$ remains unaffected; see inset of Fig.~\ref{fig:1}b.
%
%
The resulting change in spin-state splitting emerges as a phase shift in the ensemble via Ramsey interferometry, where the ensemble spins are brought into a superposition between the states $\ket{0}$ and $\ket{1}$ before Rydberg dressing is activated; see Fig.~\ref{fig:1}c.
For the control spin in $\ket{g}_C$, the target spins $\ket{0}_T$ evolve as $\sin^2(\phi/2)\ket{0}_T + \cos^2(\phi/2)\ket{1}_T$; see blue line in Fig.~\ref{fig:1}d.
Here, the indices $C$ and $T$ label the states of the control and target spins, respectively, and $\phi = \delta_{g}t_T$ equals the accumulated phase.
In contrast, when the control spin is excited to the Rydberg state $\ket{S}$, the strong $\ket{S}$-$\ket{P}$ interaction $U_{\mathrm{int}}$ renders the dressing transition substantially off-resonant for the target spins within an interaction range set by the control-target interactions.
As a consequence, the AC-Stark shift is strongly reduced to $\delta_S \ll \delta_g$, see inset Fig.~\ref{fig:1}b, and the Ramsey phase becomes negligible; see red line in Fig.~\ref{fig:1}d.
The resulting sharp change in fringe contrast directly encodes the control-spin state, providing the basis for coherent state transfer between a single qubit and the ensemble.
%
%
Here, Rydberg dressing plays two complementary roles.
First, it generates strong, switchable interactions between the control spin and the target ensemble, enabling readout of the control state.
Second, the Rydberg state admixture $\beta = \Omega_P/2\Delta$ controls the residual fourth-order intra-ensemble interactions, scaling as $\sim \beta^4 \Delta$ as illustrated in the inset of Fig.~\ref{fig:1}a, and can be tuned small so that the targets behave as independent spectator spins for the protocols presented here.
%
%

%
%
%
Our experiment begins with a near unity-filled two-dimensional array of $^{87}\mathrm{Rb}$ atoms trapped in the motional ground state of a three-dimensional optical lattice.
The interatomic spacing in the array plane is set by our folded lattice geometry~\cite{Wei2023} to $a_\mathrm{lat} = \SI{752}{\nano\meter}$, see Fig.~\ref{fig:1}a.
Using single-site addressing~\cite{Weitenberg2011, Srakaew2023}, we deterministically prepare a $3\times3$ array of atoms.
A bias magnetic field of $B = \SI{30.32}{G}$ along the vertical direction defines the quantization axis.
We initialize the central control spin in the electronic ground state manifold and hyperfine state $\ket{g} = \ket{1} = \ket{F=2,m_F = -2}$ and the surrounding target spins in the state $\ket{0} = \ket{F = 1, m_F = -1}$, where $F$ denotes the total angular momentum quantum number and $m_F$ its projection onto the quantization axis.
The transition between $\ket{0}$ and $\ket{1}$ is coherently driven via a two-photon Raman transition with a Rabi frequency of $\Omega_R = 2\pi\times\SI{1.26(1)}{\mega\hertz}$~\cite{Levine2022,Srakaew2024}.
We resonantly couple the control spin in state $\ket{g}$ to the Rydberg state $\ket{S} = \ket{36S_{1/2}, m_J = -1/2}$ with a two-photon Rabi frequency of $2\pi\times\SI{4.45(6)}{\mega\hertz}$.
We off-resonantly dress the target spins in the state $\ket{0}$ on a single-photon transition to the Rydberg states $\ket{P_{+}} = \ket{36P_{1/2}, m_J = 1/2}$ and $\ket{P_{-}} = \ket{36P_{1/2}, m_J = -1/2}$ with Rabi couplings $\Omega_{P_+} = \Omega_{P_{-}}/\sqrt{3}$ and $\Omega_{P_-} = 2\pi\times\SI{7.11(1)}{\mega\hertz}$, and detunings $\Delta_{+} = 2\pi\times\SI{11.88}{\mega\hertz}$ and $\Delta_{-} = 2\pi\times\SI{40}{\mega\hertz}$ respectively; see~\cite{SI}.
This off-resonant coupling induces an AC-Stark shift, $\delta_g$, on the $\ket{0}$ state, that can be controllably switched to $\delta_S \ll \delta_g$ by the state of the control spin.
%

%
%
%

\begin{figure}
    \centering
    \includegraphics[width=0.5\textwidth]{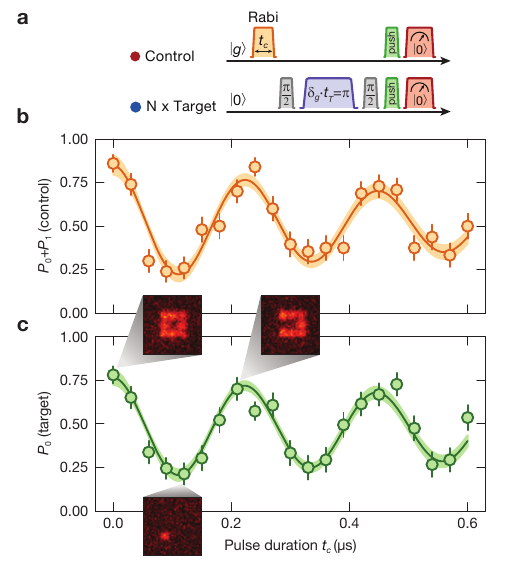}
    \caption{\textbf{Coherent control of the imprinted phase.}
    (\textbf{a}) Schematic of the protocol.
    We apply a coherent Rabi drive $\ket{g} \leftrightarrow \ket{S}$ on the control spin followed by a Ramsey measurement on the target spins with a constant duration of $\delta_g t_T = \pi$.
    This sequence maps the quantum state of the control spin onto the target spins.
    (\textbf{b}) Coherent Rabi oscillations of the control spin measured via Rydberg-induced atom loss.
    (\textbf{c}) Oscillation of the target-spin hyperfine population, showing dynamics correlated with (\textbf{b}).
    Three representative single-shot fluorescence images highlight the clear oscillatory behavior in the target spins.
    }
    \label{fig:3}
\end{figure}
In order to convert the AC-Stark shift into a time-dependent phase, we employ a Ramsey interferometry sequence on the target ensemble, as illustrated in Fig.~\ref{fig:1}c.
Applying a variable time delay between the Ramsey $\pi/2$-pulses generates an oscillation between the hyperfine states $\ket{0}$ and $\ket{1}$ of each target spin, which we detect through a spin-selective push-out of state $\ket{1}$ while measuring the remaining fraction in $\ket{0}$; see Fig.~\ref{fig:2}a.
%
%
%
From the resulting oscillation, we extract an AC-Stark shift of $\delta_{g} = 2\pi\times\SI{578(4)}{\kilo\hertz}$ in reasonable agreement with the expected shift of $\delta^{\mathrm{theory}}_{g} = 2\pi\times\SI{670}{\kilo\hertz}$. 
%
%
In contrast, the accumulated phase changes dramatically when exciting the control spin to the Rydberg $\ket{S}$ state.
%
At close control-target distances, the effective AC-Stark shift is reduced to $\delta_S = \Omega_{P_-}^2 / 4(\Delta_{-}-U^{-}_{\mathrm{int}}(r=a_{\mathrm{lat}}))+\Omega_{P_+}^2 / 4(\Delta_{+}-U^{+}_{\mathrm{int}}(r=a_{\mathrm{lat}})) \approx 2\pi\times\SI{9.8}{\kilo\hertz}$, with $U^{+(-)}_{\mathrm{int}}$ representing the $\ket{S}-\ket{P_{+(-)}}$ Rydberg interaction~\cite{SI}.
As $\delta_S$ is significantly smaller than $\delta_g$, the accumulated phase of the target spins remains negligible for probe times below $\SI{3}{\micro\second}$; see Fig.~\ref{fig:1}d.
In our experiment, however, we observe residual oscillations; see red data points in Fig.~\ref{fig:2}a.
We attribute these oscillations to state-preparation infidelities of the control spin into the $\ket{S}_C$ state; see~\cite{SI} for a detailed discussion.
Note that the phase shift depends sensitively on the $\ket{S}-\ket{P_{+(-)}}$ Rydberg interaction strength, scaling as $\propto1/r^3$ with the control-target separation $r$ at larger distances.
To probe this dependence, we fix the Ramsey phase at $\phi = \pi$, which yields maximum contrast at $r=a_{\mathrm{lat}}$; see Fig.~\ref{fig:2}b.
We observe a monotonically increasing population in $\ket{0}$ with increasing distance, consistent with the decreasing Rydberg interaction strength.
Beyond approximately four lattice sites, the Rydberg interaction is negligible relative to the dressing detuning, so the target-spin response becomes independent of the control state; see Fig.~\ref{fig:2}c.
Our experimental data show good agreement with the theoretical distance-dependence given by the ab-initio calculated Rydberg pair potentials shown in Fig.~\ref{fig:s2}a.
%

%
%
%
%
\label{sec:Atom_number_resilience}
Rydberg dressing enables a controllable interaction between the target spins and the Rydberg-excited control spin.
At the same time, at small dressing admixture $\beta$, higher-order two- and many-body interactions~\cite{Bouchoule2002,Henkel2010,Jau2016,Zeiher2016} between the target spins can be suppressed.
This feature renders our scheme intrinsically robust to variations in the target-spin number, enabling a controlled, loss-resilient phase shift.
To test this robustness, we compare the acquired phase shift for $8$ and $4$ target spins, where the four diagonal target spins are discarded; see Fig.~\ref{fig:2}c. 
The corresponding control-induced precession rates of the target spins, $2\pi\times\SI{564(4)}{\kilo\hertz}$ (for $N=8$) and $2\pi\times\SI{569(6)}{\kilo\hertz}$ (for $N=4$), agree within fit uncertainties.
This demonstrates that the controlled-phase dynamics are consistent across the two target-spin configurations and with single-particle theoretical estimates, see~\cite{SI}.
%
%
%
As a consequence of this intrinsic robustness to target spin-number fluctuations, the approach is also a candidate building block for scalable multi-qubit phase gates~\mbox{\cite{Su2018,Han2020}}.

%
%
The coherent nature of the coupling between control spin and target ensemble allows us to extend the protocol to superpositions of the control state.
To verify this experimentally, we drive coherent Rabi oscillations on the control spin followed by a Ramsey sequence on the target spins; see Fig.~\ref{fig:3}.
To maximize the expected contrast, we fix the acquired Ramsey phase to $\phi = \pi$ measured when the control qubit is in $\ket{g}$.
First, we measure coherent Rabi oscillations on the control atom by detecting Rydberg-induced atom loss, yielding a Rabi frequency of $\Omega_S = 2\pi\times\SI{4.45(6)}{\mega\hertz}$; see Fig.~\ref{fig:3}b.
Subsequently, by tracking the collective spin of the target atoms as a function of the drive duration on the control spin, we observe an oscillation with frequency $2\pi\times\SI{4.48(5)}{\mega\hertz}$, consistent with the dynamics observed for the control spin.
%
%
%
This demonstrates that the hyperfine state of the target ensemble can be coherently controlled through the control spin.
In principle, the underlying unitary interaction can be used to prepare the maximally entangled GHZ-like state $1/\sqrt{2}(\ket{g}_C\ket{0}_T^{\otimes N}+\ket{S}_C\ket{1}_T^{\otimes N})$.
Verifying this entanglement is beyond the scope of the present work: it requires a direct measurement of the correlations and coherence between the control and the target spins, which our current protocol does not allow.
In the present implementation, the spin-selective push-out used to readout the target spins precludes simultaneous detection of both subsystems.
This limitation can be overcome in future experiments by selecting a different set of hyperfine $m_F$ levels, applying non-destructive spin-resolved detection, or performing a final $\pi$-rotation on the control spin before the global push-out pulse.

%
%
%

%
\begin{figure}[t!]
    \centering
    \includegraphics[width=0.5\textwidth]{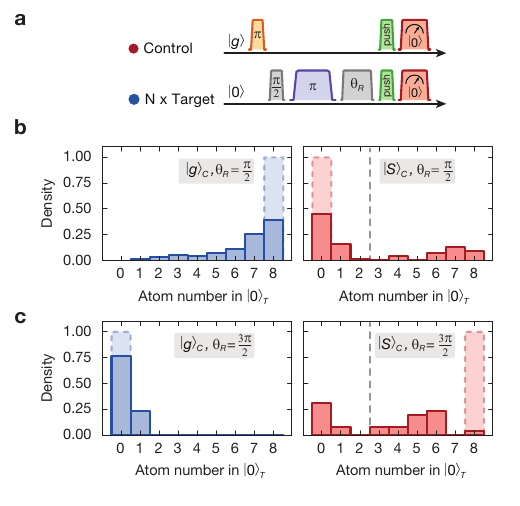}
    \caption{\textbf{Single Rydberg qubit detection.}
    (\textbf{a}) Schematic of the protocol.
    The control spin is prepared in the ground state ($\ket{g}_C$) or the Rydberg state ($\ket{S}_C$) by a $\pi$ pulse.
    The Raman pulse $\theta_R$ maps the accumulated phase onto population before measurement; $\theta_R=3\pi/2$ inverts the detection outcome compared to $\theta_R=\pi/2$.
    (\textbf{b}) Spin number histogram for $\theta_R = \pi/2$.
    For $\ket{g}_C$ (blue), the histogram is concentrated in a range of $6$ - $8$ spins.
    For $\ket{S}_C$ (red), successful Rydberg preparation results in $0$ - $1$ spin, while the $6$ - $8$ spin peak arises from imperfections in the Rydberg preparation.
    The dashed histograms represent the expected distribution under ideal conditions.
    Vertical dashed lines indicate the threshold for Rydberg spin detection.
    (\textbf{c}) Same as (b) but for $\theta_R=3\pi/2$.
    For $\ket{g}_C$ (blue), both signal and target-spin preparation error are confined to $0$ - $1$ spin, yielding a background-free region; events with $\geq 3$ detected spins unambiguously indicate a Rydberg excitation on the control spin (red).
    }
    \label{fig:4}
\end{figure}
A first application of the controlled-phase imprint is non-destructive detection of the Rydberg-encoded control spin via the target ensemble.
The key idea is to correlate the number of detected target spins with the control-spin state, exploiting the $\sqrt{N}$ enhancement of the signal-to-noise ratio with the number of target atoms.
%
%
We tune the Ramsey phase so that in the ideal case the target ensemble ends up in $\ket{0}_T^{\otimes N}$ ($\ket{1}_T^{\otimes N}$) when the control is in the ground state $\ket{g}_C$ (Rydberg state $\ket{S}_C$).
In practice, however, we observe deviations from this ideal scenario (see Fig.~\ref{fig:4}b).
When preparing the control spin in the ground state $\ket{g}_C$, the histogram peaks at high spin numbers but exhibits a pronounced tail towards a lower number of $\ket{0}_T$ spins due to fluctuations in the initially prepared number of target spins, e.g, due to defects in the array.
In contrast, when the control spin is prepared in the Rydberg state $\ket{S}_C$, the distribution peaks around zero spins, with a tail towards higher counts due to imperfect preparation of the control spin in the Rydberg state.
Taking into account the overlap of the histograms, we estimate an upper bound on the control-spin state-assignment fidelity of $\mathcal{F} \lesssim \SI{93}{\percent}$~\cite{SI}.
To further improve the Rydberg spin detection fidelity, we benchmark a modified Ramsey sequence in which we apply an additional phase of $\pi$ before the final pulse, inverting the detection logic; see Fig.~\ref{fig:4}c.
This modification enables effectively background-free detection of the Rydberg excitation, signaled by a number of target spins in $\ket{0}_T$ exceeding a threshold --- in analogy to an avalanche detector for single photons.
The effective absence of background arises because other events, including correctly prepared ground-state spins ($\ket{g}_C$), loss of the control spin, and Rydberg-state-preparation errors, yield zero or one detected target spins.
%
%
%
Setting the Rydberg-detection threshold at $\geq 3$ target spins yields a Rydberg-state assignment fidelity of $\mathcal{F} = 99.81^{+0.17}_{-1.47} \si{\percent}$ using Bayes' theorem; see~\cite{SI} for details.
In principle, the spin-number insensitivity of our protocol allows us to further increase the number of target spins without degrading the per-atom performance, with a corresponding $\sqrt{N}$ improvement in distinguishing the two histograms.
%

%
%
%
%

%
Our work demonstrates a new mechanism to coherently couple a single Rydberg excitation to a Rydberg-dressed spin ensemble.
Leveraging interaction control by the Rydberg admixture, we suppress interactions within the ensemble, making the scheme robust to fluctuations in the number of target spins.
This approach paves the way for high-fidelity Rydberg-state readout and faster Rydberg detection through increased photon collection rates from the target ensemble~\cite{Petrosyan2024}.
The collocation of such target ensembles with Rydberg spins furthermore enables non-destructive measurement of Rydberg excitations in a quantum simulator.
Additional opportunities emerge from controlling the quantum-optical response of sub-wavelength target arrays~\cite{Asenjo2017, Rui2020, Srakaew2023}.
Furthermore, optical cavities could be used to efficiently read out the state of the collective target ensemble~\cite{Deist2022, Hu2025, DeSantis2026}.
Our current single-species implementation is limited by crosstalk between the addressing of control and target spins, which share a common spin basis.
In future experiments, the target spins could be encoded in the hyperfine clock states ($\ket{F=1, m_F=0}$ and $\ket{F=2, m_F=0}$) of $^{87}\mathrm{Rb}$, realizing a dual-qubit architecture with independent control of both qubits.
The enhanced lifetime and coherence time of these ground-state qubits will enable fast mid-circuit readout, the deterministic preparation of multi-atom entangled states or the implementation of advanced multi-qubit gates, such as the $\mathrm{C}_k\mathrm{NOT}$ gate~\cite{Isenhower2011,Petrosyan2024, delakouras2025,Vaknin2026}.
Our results establish a platform for non-destructive Rydberg readout with concrete perspectives for quantum simulation, quantum information processing, and quantum optics with neutral atoms.

\begin{acknowledgments}

\textit{Acknowledgment:} We thank Guillaume Brochier for providing feedback on the manuscript.
We acknowledge funding by the Max Planck Society (MPG) and the Deutsche Forschungsgemeinschaft (DFG, German Research Foundation) under Germany's Excellence Strategy--EXC-2111--390814868, and from the Munich Quantum Valley initiative as part of the High-Tech Agenda Plus of the Bavarian State Government.
We also acknowledge funding through JST-DFG 2024: Japanese-German Joint Call for Proposals on ``Quantum Technologies'' (Japan-JST-DFG-ASPIRE 2024) under DFG Grant No. 554561799.
This publication has also received funding under the Horizon Europe program HORIZON-CL4-2022-QUANTUM-02-SGA via the project 101113690 (PASQuanS2.1).
P.W. acknowledges funding through the Walter Benjamin program (DFG project 516136618).
J.Z. acknowledges support from the BMFTR through the program ``Quantum technologies - from basic research to market'' (Grant No. 13N16265).
K.S. and S.A. acknowledge funding from the International Max Planck Research School (IMPRS) for Quantum Science and Technology.
D.G. acknowledges funding from the Munich Quantum Valley initiative as part of the High-Tech Agenda Plus of the Bavarian State Government.

\end{acknowledgments}


\bibliography{SPgate}
\clearpage


\setcounter{equation}{0}
\setcounter{figure}{0}
\setcounter{table}{0}
\renewcommand{\theequation}{S\arabic{equation}}
\renewcommand{\thefigure}{S\arabic{figure}}
\renewcommand{\thetable}{S\arabic{table}}

\section*{Supplementary Information}

\section{Initial state preparation and electronic level scheme}
\label{Initial_preparation}
Here we provide details of the initial state preparation, the relevant atomic levels, and the excitation scheme used in the experiment.
We begin by preparing a nearly unity-filled two-dimensional atomic array of around $100$ atoms in the hyperfine ground state $\ket{0} = \ket{F = 1, m_F = -1}$ trapped in an optical lattice with a lattice spacing of $a_{\mathrm{lat}} = \SI{752}{\nano\meter}$.
Using single-site addressing~\cite{Weitenberg2011, Srakaew2023}, the initial state is prepared in two steps.
First, a $3\times 3$ sparse atomic array is prepared in $\ket{0}$ with a success fidelity of $0.94(1)$ at interatomic spacing $r = a_{\mathrm{lat}}$ for most experiments presented in the main text, or $r = 2a_{\mathrm{lat}}$ and $3a_{\mathrm{lat}}$ for the experiment shown in Fig.~\ref{fig:2}b.
%
%
Subsequently, the control spin at the center of the array is transferred to the hyperfine state $\ket{g} = \ket{F=2,m_F = -2}$ with a fidelity of $0.87(1)$, resulting in an initial configuration consisting of a single control spin in $\ket{g}$ surrounded by an ensemble with eight target spins in $\ket{0}$.

\begin{figure}[!]
    \centering
    \includegraphics[width=0.5\textwidth]{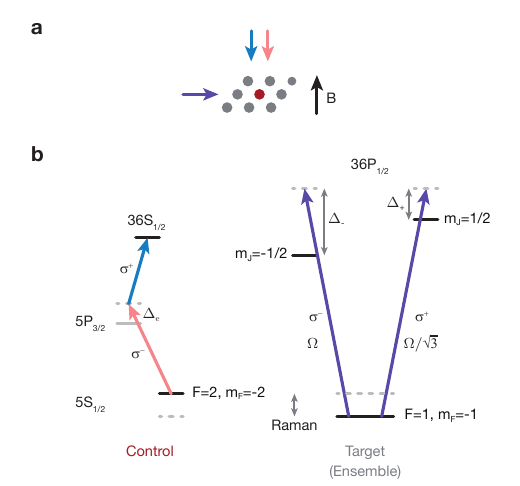}
    \caption{\textbf{Electronic level and excitation scheme.}
    (\textbf{a}) The initial atomic states and beam configuration.
    A magnetic field is applied orthogonal to the atomic plane.
    The $\SI{780}{\nano\meter}$ (red arrow) and $\SI{480}{\nano\meter}$ (blue arrow) beams co-propagate parallel to the magnetic field, enabling pure $\sigma^-$ and $\sigma^+$ polarizations, respectively.
    The $\SI{297}{\nano\meter}$ (purple arrow) beam propagates within the atomic plane with linear polarization orthogonal to the magnetic field, resulting in coupling to the Rydberg manifold via both $\sigma^{-}$ and $\sigma^{+}$ transitions.
    (\textbf{b}) Excitation scheme.
    The control spin is excited to the Rydberg state $\ket{S}=\ket{36S_{1/2},m_J=-1/2}$ via a two-photon transition with an intermediate-state detuning of $\Delta_e = 2\pi\times\SI{200}{\mega\hertz}$.
    The target spins are off-resonantly coupled to the $36P_{1/2}$ Rydberg states with detunings of $\Delta_{-} = 2\pi\times\SI{40}{\mega\hertz}$ and $\Delta_{+} = 2\pi\times\SI{11.88}{\mega\hertz}$ for the $\sigma^-$ and $\sigma^+$ transitions, respectively.
    }
    \label{fig:s1}
\end{figure}

Coherent transfer between the states $\ket{1}$ and $\ket{0}$ is achieved using a fast two-photon Raman transition in a $\sigma$–$\pi$ polarization configuration~\cite{Srakaew2024}.
The Raman beam globally addresses the atomic array with a two-photon Rabi frequency of $\Omega_R = 2\pi\times\SI{1.26(1)}{\mega\hertz}$, achieving a $\pi$-pulse population transfer efficiency exceeding $99\,\%$.
For the protocol discussed in the main text, it is essential that the population transfer between $\ket{0}$ and $\ket{1}$ can be performed faster than the lifetime of the Rydberg state $\ket{S}$.
Our excitation scheme to the Rydberg states is illustrated in Fig.~\ref{fig:s1}b.
The control spin in the state $\ket{g}$ is excited to the state $\ket{S}=\ket{36S_{1/2},m_J=-1/2}$ via the intermediate state $\ket{e}=\ket{5P_{3/2},F=3,m_F=-3}$.
This two-photon excitation employs $\sigma^{-}$-polarized light at a wavelength of $\SI{780}{\nano\meter}$ and $\sigma^{+}$-polarized light at $\SI{480}{\nano\meter}$.
With an intermediate state detuning of $\Delta_e=2\pi\times\SI{200}{\mega\hertz}$, we obtain a two-photon Rabi frequency of $\Omega_S = 2\pi\times\SI{4.45(6)}{\mega\hertz}$ with an exponential damping time of $\SI{0.9(4)}{\micro\second}$.
This results in a Rydberg excitation fidelity of $0.94(3)$ for a $\pi$-pulse.
Combined with the $3\times3$ array preparation fidelity, the total probability to prepare a single control spin in $\ket{S}$ is $P_{\mathrm{init}}=0.82(3)$.
The target spins prepared in $\ket{0}$ are off-resonantly coupled to the Rydberg states $\ket{P_{-}} = \ket{36P_{1/2}, m_J = -1/2}$ and $\ket{P_{+}} = \ket{36P_{1/2}, m_J = 1/2}$ using a single-photon excitation at a wavelength of $\SI{297}{\nano\meter}$.
The detunings from $\ket{P_{-}}$ and $\ket{P_{+}}$ are $\Delta_{-} = 2\pi\times\SI{40}{\mega\hertz}$ and $\Delta_{+} =2\pi\times\SI{11.88}{\mega\hertz}$, respectively.
The corresponding Rabi frequencies are $\Omega_{P_{-}} = 2\pi\times\SI{7.11(1)}{\mega\hertz}$ and $\Omega_{P_{+}} = \Omega_{P_{-}}/\sqrt{3}$.

\section{Pair-wise interaction}
\begin{figure}[!]
    \centering
    \includegraphics[width=0.5\textwidth]{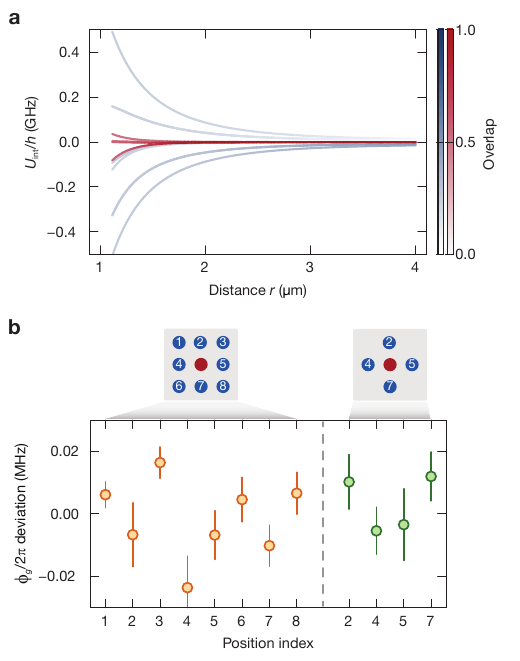}
    \caption{\textbf{Rydberg interactions and individual AC-Stark shifts of all target spins.}
    (\textbf{a}) Rydberg interaction strengths $U_{\mathrm{int}}$ obtained using the ``pairinteraction'' package~\cite{Weber2017}.
    The $\ket{P_i} - \ket{P_i}$ interactions (red) are much weaker than the $\ket{S} - \ket{P_i}$ interactions (blue).
    For the $\ket{S} - \ket{P_{+(-)}}$ interaction, the coefficients are $C^{+(-)}_3/h = -354(-680)\si{\mega\hertz\cdot\micro\meter^3}$ and $C^{+(-)}_6/h = -128(-25)\si{\mega\hertz\cdot\micro\meter^6}$, yielding $U_{\mathrm{int}}^{+(-)}(r=a_{\mathrm{lat}})/h = -1.54(-1.74)\si{\giga\hertz}$.
    (\textbf{b}) Spatial variation of the AC-Stark shift $\delta_g$ across individual target spins.
    For configurations with $8$ (orange) and $4$ (green) target spin number, the deviations from the mean are approximately $\SI{3}{\percent}$.
    Insets label the position index of each target spin.
    }
    \label{fig:s2}
\end{figure}

In this section, we analyze the relevant Rydberg interactions and justify the effective pair-wise interaction used in the main text.
The AC-Stark shift acquired by the target spin is given by
\begin{equation}
\delta_g = \sum_i \left( \frac{\Omega_{P_i}^2}{4\Delta_i}-\frac{\Omega_{P_i}^4}{8\Delta_i^3}+\sum_j\frac{\Omega_{P_i}^4}{8\Delta_i^2(2\Delta_i-U_{\mathrm{int}}^{\ket{P_i} - \ket{P_i}}(r_j))} \right),
\end{equation}
where $i=\in\{+,-\}$ labels the two Rydberg P-states and $j$ labels the target spin index~\cite{Srakaew2024,Zeiher2017}.
The first term is the standard second-order AC-Stark shift in the dressing limit, while the latter two terms arise from fourth-order processes that incorporate the distance dependent intra-target Rydberg interactions $U_{\mathrm{int}}^{\ket{P_i}-\ket{P_i}}(r_j)$.
In our experiment, the parameters are chosen such that the second-order contribution dominates.
The higher-order corrections amount to only $\sim\SI{2}{\percent}$ of the total phase shift, validating a perturbative treatment.
Furthermore, the Rydberg interaction between the control and target spins, $\ket{S}-\ket{P_i}$ (blue curves in Fig.~\ref{fig:s2}a), is substantially stronger than the interaction between target spins, $\ket{P_i}-\ket{P_i}$ (red curves).
Consequently, many-body effects among the target spins are negligible (see Fig.~\ref{fig:s2}b), and the dynamics are well described by independent pair-wise interactions between the control spin and each target spin individually.
The AC-Stark shift $\delta_g$ must also be sufficiently large that significant evolution phases can be picked up at timescales much shorter than the lifetime of the control-spin Rydberg state.
When the control spin is excited to $\ket{S}$, the Rydberg interaction shifts the target-spin Rydberg states and suppresses the second-order AC-Stark shift.
In this case, the AC-Stark shift simplifies to
\begin{equation}
\delta_S = \sum_i \Omega_{P_i}^2 / 4(\Delta_{i}-U^{\ket{S} - \ket{P_i}}_{\mathrm{int}}),
\end{equation}
where the control spin interacts independently with each target spin.

\section{Atomic loss from off-resonant coupling}
\begin{figure}[!]
    \centering
    \includegraphics[width=0.5\textwidth]{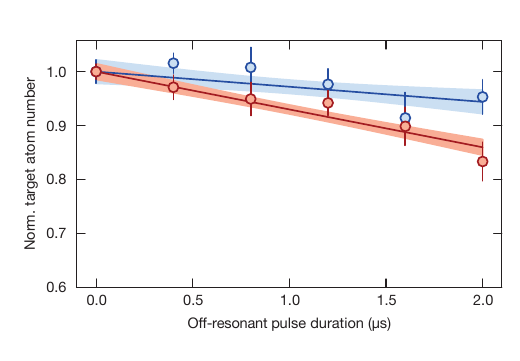}
    \caption{\textbf{Atom loss in the Ramsey protocol.}
    We observe the target spin loss due to the off-resonant coupling to the Rydberg state $36P_{1/2}$.
    The blue (red) plot shows the case where the control spin is being prepared in the ground state $\ket{g}$ (Rydberg state $\ket{S}$).
    }
    \label{fig:s3}
\end{figure}

We investigate the atomic loss of the target spins arising from off-resonant coupling to $\ket{P_i}$ with the protocol illustrated in Fig.~\ref{fig:1}c. 
The protocol is modified to enable readout of both $\ket{0}$ and $\ket{1}$ populations by omitting the $\ket{1}$ resonant push-out pulse.
As shown in Fig.~\ref{fig:s3}, we measure an atomic loss rate of $0.03(2)$ atoms/$\SI{}{\micro\second}$ (averaged over the eight-atom ensemble) when the control spin is prepared in the ground state $\ket{g}$.
In contrast, when the control spin is prepared in the Rydberg $\ket{S}$ state, the loss rate increases to $0.07(1)$ atoms/$\SI{}{\micro\second}$.
This enhanced loss is attributed to Rydberg interactions between $\ket{S}$ and the $\ket{P_-}$ and $\ket{P_+}$ states, which can open additional decay channels~\cite{Zeiher2016, Festa2022}.

\section{Deriving and estimating the detection fidelities $\mathcal{F}$}
In the following, we derive the fidelity for detecting Rydberg states using the two schemes presented in the main text (see Fig.~\ref{fig:4}).
In the first sequence, we apply only a final $\theta_R=\pi/2$ rotation in the Ramsey sequence (see Fig.~\ref{fig:4}b).
In this case, we expect all target atoms to be present (absent) in $\ket{0}_T$ when the control atom is prepared in the ground state $\ket{g}_C$ (Rydberg state $\ket{S}_C$).
To quantify a state assignment fidelity $\mathcal{F}$, we make use of Youden's J statistic~\cite{youden1950index} given by
\begin{equation}
    \mathcal{F} = \frac{P(g_c|g_c)}{P(g_c|g_c)+P(S_c|g_c)} + \frac{P(S_c|S_c)}{P(S_c|S_c)+P(g_c|S_c)} - 1.
\end{equation}
Here $P(X|Y)$ are the entries of the confusion matrix, representing the likelihood of assigning state $X$ when state $Y$ was prepared.
We estimate the entries of $P(X|Y)$ by evaluating the histograms in Fig.~\ref{fig:4}.
We set the assignment threshold at $3$ target spins: events with $\geq 3$ spins are assigned to the ground state, while  all other events are assigned to the Rydberg state.
Given our preparation fidelity in Fig.~\ref{fig:4}b, we find the true positive $P(g_c|g_c)\sim \SI{94}{\percent}$ and the false negative $P(S_c|g_c)\sim \SI{6}{\percent}$.
Determining the corresponding confusion-matrix entries for initial preparation in $\ket{S}_C$ is more challenging because we cannot reliably know on a single-shot basis whether the Rydberg atom was actually prepared.
Nevertheless, to first order, the influence of the Rydberg atom can be estimated from the histogram for the second sequence with $\theta_R=3\pi/2$, since the presence of a Rydberg atom and a longer Raman pulse both invert the detection logic.
From the histogram, we conservatively estimate $P(S_c|S_c)\lesssim \SI{99}{\percent}$ and $P(g_c|S_c)\gtrsim \SI{1}{\percent}$.
Combining the confusion-matrix elements gives a state-assignment fidelity of $\mathcal{F}\lesssim \SI{93}{\percent}$ for the first Ramsey sequence.
The second Ramsey sequence under discussion closes with a final $\theta_R=3\pi/2$ rotation (see Fig.~\ref{fig:4}c).
In this configuration, preparation errors become indistinguishable from the ground-state case $\ket{g}_C$, both resulting in low-count events on the target spins; see left histogram in Fig.~\ref{fig:4}c.
In the absence of Rydberg excitations, each target spin behaves as an independent Bernoulli trial, in analogy with the single-photon amplification mechanism in an avalanche photodetector.
For independent events, we can derive a confidence interval for the underlying binomial distribution, known as the Wilson score interval~\cite{Wilson1927}
\begin{equation}
    \frac{1}{1+\frac{z^2}{n}} \left[ p + \frac{z^2}{2n} \pm z\sqrt{\frac{p(1-p)}{n} + \frac{z^2}{4n^2}}  \right].
\end{equation}
Here, $z=1.96$ is the standard normal quantile for a $\SI{95}{\percent}$ confidence interval, $p$ is the per-trial success probability estimate, and $n$ is the total number of independent observations.
In our case, we accumulated $26$ shots with up to $8$ target spins each, yielding up to $208$ independent events from which we extract a per-target fidelity of $97.1^{+1.6}_{-3.3}\si{\percent}$.
The asymmetric error bar reflects the limited number of shots; tightening the lower bound requires only modest additional statistics.
Following the binomial distribution, we can estimate the probability of measuring at least three target spins, if initially preparing the control atom in its ground state, yielding $P(\geq 3\ \mathrm{cts}|g_c)\approx 0.12^{+0.91}_{-0.11}\si{\percent}$.
For our observed histogram, we can also derive the probability to count at least three target atoms when preparing the control atom in the Rydberg state, yielding $P(\geq 3\ \mathrm{cts}|S_c)\approx 61.5\si{\percent}$.
With these two quantities, we can derive the Rydberg-detection fidelity using Bayes' theorem
\begin{equation}
    P(S_c|\geq 3\ \mathrm{cts}) = \frac{P(\geq 3\ \mathrm{cts}|S_c)\cdot P(S_C)}{P(\geq 3\ \mathrm{cts})}.
\end{equation}
Note, that $P(\geq 3\ \mathrm{cts}) = P(\geq 3\ \mathrm{cts}|S_c)\cdot P(S_C) + P(\geq 3\ \mathrm{cts}|g_c)\cdot P(g_C)$ and $P(S_C) = P(g_C) = 0.5$ for standard experiments.
Altogether, this yields a Rydberg assignment fidelity of $P(S_c|\geq 3\ \mathrm{cts}) = 99.81^{+0.17}_{-1.47}\si{\percent}$, when choosing a threshold of at least three detected target spins in $\ket{0}$.

\clearpage

\end{document}